\documentclass[pre,amssymb,reprint,superscriptaddress,showpacs]{revtex4-1}
\usepackage{amsmath}
\usepackage{amsfonts}
\usepackage{amssymb}
\usepackage{epsfig}
\usepackage{graphicx}
\newcommand{\gdot}[0]{\dot{\gamma}}
\newcommand{\Pe}{P_{\rm e}}
\renewcommand{\phi}{\varphi}

\newcommand{\be}{\begin{equation}}
\newcommand{\ee}{\end{equation}}
\newcommand{\bea}{\begin{equnaray}}
\newcommand{\eea}{\end{equnaray}}
\newcommand{\ba}{\begin{align}}
\newcommand{\ea}{\end{align}}

\usepackage{color}
\usepackage{ulem}

\begin{document}

\title{Discontinuous shear-thickening in Brownian suspensions} 
 
\author{Takeshi Kawasaki}

\affiliation{Department of Physics, Nagoya University, Nagoya 464-8602, Japan}

\author{Ludovic Berthier}
\affiliation{Laboratoire Charles Coulomb (L2C), University of Montpellier, CNRS, Montpellier, France}

\date{\today}

\begin{abstract}
Discontinuous shear-thickening in dense suspensions naturally emerges from the activation of frictional forces by shear flow in non-Brownian systems close to jamming. Yet, this physical picture is incomplete as most experiments study soft colloidal particles subject to thermal fluctuations. To characterise  discontinuous shear-thickening in colloidal suspensions we use computer simulations to provide a complete description of the competition between athermal jamming, frictional forces, thermal motion, particle softness, and shear flow. We intentionally neglect hydrodynamics, electrostatics, lubrication, and inertia, but can nevertheless achieve quantitative agreement with experimental findings. In particular, shear-thickening corresponds to a crossover between frictionless and frictional jamming regimes which is controlled by thermal fluctuations and particle softness and occurs at a softness dependent P\'eclet number. We also explore the consequences of our findings for constant pressure experiments, and critically discuss the reported emergence of `S-shaped' flow curves. Our work provides the minimal ingredients to quantitatively interpret a large body of experimental work on discontinuous shear-thickening in colloidal suspensions.
\end{abstract}



\maketitle
\section{Introduction}

The flow behavior of industrially-relevant complex fluids is often non-linear, with flow curves typically exhibiting multiple rheological regimes, changing between Newtonian, shear-thinning and shear-thickening behaviours depending on external control parameters~\cite{Larson_book,wagner_book,wagner_physicstoday}. A central goal in rheology is to make sense of these various regimes and to disentangle the competing physical origins of these nonlinearities in order to design materials with desired rheological properties. Whereas shear-thinning typically stems from the `disruption' of the material micro-structure by the shear flow, shear-thickening is interpreted as the opposite trend of a shear flow that `enhances' some underlying structural organisation~\cite{Larson_book,wagner_book}. However, because an imposed shear flow represents an external forcing, thickening is less common and thus more difficult to understand than thinning. Shear-thickening is nevertheless observed experimentally in many types of suspensions~\cite{Larson_book,wagner_book,wagner_physicstoday,WagnerJCP,fall08,brown_review,brown,fernandez,poon} that span a broad range of particle sizes, shapes, and interparticle interactions. The fact that it is a fun phenomenon observable in everyday life adds to its appeal as an object of academic study.    

A smooth and continuous increase of the shear viscosity upon increasing the shear rate may be explained by several physical processes, from hydrodynamic to inertial effects~\cite{wagner_physicstoday,brown_review,claudin,kawasaki14,otsubo,fall}. A more severe form of thickening is the discontinuous upward jump of the viscosity typically observed in denser suspensions. A recent flurry of experimental~\cite{royer,solomon,clavaud,lyderic1,lyderic2,isa} and theoretical~\cite{otsuki11,seto13,wyart,heussinger,mari14,mari15s,mari15,heussinger15,mari17,trulsson,seto17} activity has convincingly established that discontinuous shear-thickening can be interpreted as the crossover from frictionless to frictional rheologies as the shear rate is increased in the limit of athermal hard particles. Close to jamming, frictional particles have a much larger viscosity than frictionless ones~\cite{silbert02,silbert10,vanhecke_review}, and a sharp mobilisation of frictional forces upon increasing the shear flow naturally accounts for discontinuous shear-thickening. Shear-thickening becomes weaker far from jamming, and other interpretations may then become possible~\cite{wagner_physicstoday}. This scenario was illustrated in specifically-designed computer experiments, and received ample experimental confirmation over the last few years. In particular, the link between frictional forces and the existence of discontinuous shear-thickening was directly demonstrated in very elegant experiments~\cite{clavaud,solomon,lyderic1,isa}, as well as several computer models~\cite{seto13,heussinger15,trulsson,seto17} that involve a variety of choices for interparticle interactions, microscopic dynamics, and various levels of realism regarding hydrodynamic flows. 

The central role played by frictional rheology relies on the idea that the material lives somehow `close' to the jamming transition observed in the limit of non-Brownian suspensions with simple repulsive interactions~\cite{vanhecke_review}. Yet, a large number of materials displaying discontinuous shear-thickening are composed of colloidal particles subject to thermal fluctuations. This is a crucial point, since the interplay between thermal forces and shear flow near glass and jamming transitions gives rise to complex rheological behaviours~\cite{ikeda,ikeda_soft}. In particular, jamming rheology is typically not pertinent for thermal colloidal particles over experimentally-relevant timescales and stress scales. Therefore, it is quite surprising that a fully non-Brownian mechanism (mobilisation of frictional forces near athermal jamming) may dictate the rheology of Brownian suspensions. Understanding why and how this may happen, and under which physical conditions discontinuous shear-thickening may arise in Brownian suspensions are the central questions tackled in the present work.

In this work we study the role of thermal fluctuations in dense suspensions of soft repulsive particles, analysing the behaviour of hard sphere suspensions as a mathematical limit within our more general model. To make quantitative progress, we intentionally make severe approximations and neglect any other form of particle interactions (such as electrostatic repulsion or particle adhesion) as well as all interactions generated by hydrodynamic forces. As a result, we are able to propose a complete quantitative understanding of the rheological behaviour of Brownian soft repulsive spheres with frictional interactions over the complete parameter space. To connect to recent experimental findings we perform simulations keeping either the volume or the confining pressure constant, shedding additional light on the role of dilatancy, potential flow instability and `S-shaped' flow curves in systems exhibiting discontinuous shear-thickening~\cite{mari15s,heussinger15,poon2,Sshaped,olmsted}. Our main finding is that our model provides a robust minimal description to understand quantitatively the various regimes of the flow curves observed in shear-thickening colloidal suspensions. In particular, we provide all details of the crossover between linear and non-linear rheologies, athermal and thermal rheologies, and the onset of shear-thickening for soft and hard particles. We suggest that any additional physical ingredient will only introduce minor quantitative changes to the behaviour reported here, and detailed work should be performed in an experiment to actually expose clear deviations from the minimal modelling analysed here.    

This article is organized as follows. 
In Sec.~\ref{model}, we present our numerical model to investigate discontinuous shear-thickening, the numerical integration for both constant volume and constant pressure simulations, and the key control parameters. 
In Sec.~\ref{macroscopic}, we present the main features of the macroscopic flow curves obtained in constant volume simulations, and derive a dynamic state diagram of the model.  
In Sec.~\ref{microscopic}, the physical origin of the discontinuous shear-thickening activated by thermal fluctuations is directly established. 
In Sec.~\ref{quantitative}, we provide quantitative comparison of our numerical model with experimental work, and discuss the hard sphere limit. 
In Sec.~\ref{pressure}, we present the results of constant pressure simulations and discuss the status of `S-shaped' flow curves. 
Finally, in Sec.~\ref{conclusions} we summarize our results and provide some perspectives. 

\section{Rheology of thermal soft spheres with friction} 

\label{model}

\subsection{Langevin dynamics for harmonic spheres with friction}

We perform overdamped Langevin dynamics simulations of a simple model for Brownian suspensions with tangential frictional contacts and simple harmonic repulsive forces~\cite{otsuki11,mari14,silbert02,silbert10}, neglecting all other types of interactions and hydrodynamic effects. In particular, tangential forces are directly related to contact forces in this model, whereas they could in principle be mediated by the background fluid in real suspensions, an effect that we can not describe, by construction.  

Our model is an equimolar binary mixture of $N$ particles with two different diameters interacting via a harmonic repulsive potential~\cite{durian}. 
The diameters of the small and large particles are 
$a$ and $1.4a$, respectively. That choice for the size dispersity is known to efficiently suppress crystallisation. The resulting volume fraction is $\varphi= \pi Na^3(1+1.4^3)/12L^3$, where $L$ is the linear dimension of the system. Periodic boundary conditions are used, and we make sure that the system remains homogeneous for all reported conditions. We perform simulations with a total number of particles $N=1000$ for most simulations, and we additionally simulate $N=10000$ particles when finite size effects need to be investigated. 

The particles evolve according to the following Langevin equations of motion 
based on the Discrete Element Method~\cite{cundall}, with thermal fluctuations. The translational motion of particle $j \in \{1,2, ... N\}$ obeys the following Langevin equation:
\be 
\sum_{k} \left[ \vec{f}_{jk}^{\rm nor}(t)+ \vec{f}_{jk}^{\rm tan}(t) \right] +\vec{f}_{j}^{\rm drag}(t)+\vec{f}_{j}^{\rm therm}(t) = \vec{0}.
\label{em}
\ee
The first and second terms of Eq.~(\ref{em}) are the interaction forces for the normal and tangential directions, respectively.
For the normal direction, the interaction force for two particles $j$ and $k$ having diameters $a_j$ and $a_k$ is a harmonic repulsion represented as
\be 
\vec{f}_{jk}^{\rm nor} = \epsilon_{\rm n}h_{jk}\vec{n}_{jk},
\label{force}
\ee 
where $\vec{n}_{jk} =\vec{r}_{jk}/|\vec{r}_{jk}|$ and $h_{jk}= (a_{jk}-|\vec{r}_{jk}|)\Theta(a_{jk}-|\vec{r}_{jk}|)$; $\Theta(x)$ is the Heaviside function, $a_{jk}=(a_{j}+a_{k})/2$, and $\vec{r}_{jk}=\vec{r}_j-\vec{r}_k$, $\vec{r}_j$ is the position of particle $j$. The tangential force is represented as $\vec{f}_{jk}^{\rm tan}=\epsilon_{\rm t} \vec{\xi}_{jk}$. Here, we follow previous work and set $\epsilon_{\rm t}=0.25\epsilon_{\rm n}$~\cite{silbert02}, while $\vec{\xi}_{jk}$ is the vector of overlap between the particles for the tangential direction defined as $\vec{\xi}_{jk}(t) = \int_{t' \in t_{\rm coll}}{\rm dt'} \vec{u}_{jk}(t')$, where $t_{\rm coll}$ is the time spent since the start of a binary collision between particles $j$ and $k$,
$\vec{u}_{jk}=(\overleftrightarrow{1}-\vec{n}_{jk}\vec{n}_{jk})\cdot\{\vec{v}_j-\vec{v}_k- \frac{1}{2}(a_j\vec{\omega}_j + a_k\vec{\omega}_k)\times 
\vec{n}_{jk}\}$. We defined $\vec{v}_{j}$ and $\vec{\omega}_{j}$ as the translational and angular velocities of particle $j$, respectively. 
{In order to make the direction of the tangential force exactly tangential, in each time step the vector of tangential overlap is updated as~\cite{luding},
\be 
\vec{\xi}_{jk}(t) - (\vec{\xi}_{jk}(t)\cdot \vec{n}_{jk})\vec{n}_{jk}  \to \vec{\xi}_{jk}(t).
\ee 
}
The tangential force has an upper bound represented as 
\be 
|\vec{f}_{jk}^{\rm tan}| \leq \mu_{\rm C} |\vec{f}_{jk}^{\rm nor}|,
\label{fric}
\ee 
which defines the friction coefficient $\mu_{\rm C}$.
When the tangential force does not satisfy Eq.~(\ref{fric}), the overlap vector for the tangential direction is scaled as $\mu_{\rm C} \frac{|\vec{f}_{jk}^{\rm nor}| }{|\vec{f}_{jk}^{\rm tan}| }\vec{\xi}_{jk} \to \vec{\xi}_{jk}$.
The third term in Eq.~(\ref{em}) is the Stokes drag force. For a shear flow in the $xy$ plane, this is written as $\vec{f}_{j}^{\rm drag}=\xi_{\rm n} \{\vec{v}_j(t)-\gdot y_j(t)\vec{e}_x \}$, 
where Lees-Edward periodic boundary conditions are implied~\cite{LeesEdwards}.
Finally, the fourth term in Eq.~(\ref{em}) is the thermal force acting on particle $j$. This thermal force is drawn from a random Gaussian distribution with zero mean and variance obeying the fluctuation-dissipation relation, $\langle \vec{f}_j^{\rm therm}(t)  \vec{f}_k^{\rm therm}(t')\rangle=2k_{\rm  B}T\xi_{\rm n}\delta_{jk} \overleftrightarrow{1} \delta(t-t')$, where $k_{\rm B}$ is the Boltzmann constant and $T$ is the temperature.

Next, we introduce the equations of motion of the particles for the rotational degrees of freedom, which are represented by
\be
\sum_{k} \vec{T}^{\rm tan}_{jk}(t)+\vec{T}^{\rm drag}_{j}(t)+\vec{T}_j^{\rm therm}=\vec{0}.
\label{rem}
\ee 
The first term is the torque of the tangential force, such that  
$\vec{T}_{jk}^{\rm tan}(t) = \vec{r}_{jk} \times \vec{f}_{jk}^{\rm tan}$.
The second term is the dissipation torque, which reads  
$\vec{T}^{\rm drag}_{j}(t)=\xi_{\rm t}(\vec{\omega}_j+\dot{\gamma}\vec{e}_z/2)$ 
where $\xi_{\rm t}=a^2\xi_{\rm n}$.
The third term is the torque exerted by the thermal force 
acting on particle $j$, which also satisfies a fluctuation-dissipation theorem, $\langle \vec{T}_j^{\rm therm}(t)  \vec{T}_k^{\rm therm}(t')\rangle = 2 k_{\rm B}T \xi_{\rm t} \delta_{jk} \overleftrightarrow{1} \delta(t-t')$.

\subsection{Constant pressure simulations}

\label{constantP}

The results shown in the first part of the paper, Secs.~\ref{macroscopic}--\ref{quantitative}, are obtained by performing simulations at constant volume, so that the volume fraction is by construction imposed during these simulations, and the pressure fluctuates freely.   

In the final part of the paper, Sec.~\ref{pressure}, 
constant pressure simulations~\cite{Feller_constPressure, kolb} are performed.
To this end, an additional equation of motion needs to be considered, as the total volume of the system now evolves dynamically with its own Langevin dynamics, 
\be 
\xi_{\rm V} \dot{V}(t)-\left[ P-\hat{P}(t) \right]  + F^{\rm therm} =0,
\label{piston}
\ee
where $P$ is the prescribed pressure, and we set $\xi_{\rm V}=10^{-4}$~\cite{kolb,kawasaki15}. The third term of Eq.~(\ref{piston}) stands for thermal fluctuations acting on the box size, and it again satisfies a fluctuation-dissipation relation given by $\langle F^{\rm therm }(t)F^{\rm therm}(t') \rangle =2 k_{\rm B} T \xi_{\rm V} \delta(t-t')$. The instantaneous value of the pressure, $\hat{P}(t)$, is defined as 
$\hat{P}(t)=\frac{1}{3V(t)}\sum_{k<j}\vec{r}_{jk} \cdot (\vec{f}_{jk}^{\rm nor}+\vec{f}_{jk}^{\rm tan})$. By construction, Eq.~(\ref{piston}) imposes that 
$\hat{P}(t)$ fluctuates around the prescribed value $P$, once steady state has been reached. In that case, the volume fraction fluctuates freely.  

\subsection{Physical units and numerical protocols}

\label{unit}

As discussed above, the repulsive, frictional, and thermal forces define the microscopic parameters that fully specify our model and its dynamics. To describe the evolution of the system we first need to define the units used for all quantities. 

Lengthscales are expressed in units of the particle diameter $a$, and timescales are expressed in units of the quantity $t_0 \equiv a^2\xi_{\rm n} /\epsilon_{\rm n}$, defined from the translational equation of motion. The microscopic timescale $t_0$ represents the typical dissipation timescale in Eq.~(\ref{em}). As a result, shear rates are expressed in units of $t_0^{-1}$. Pressures and shear stresses are expressed in units of 
\be 
\sigma_0 = \epsilon_{\rm n}/a^3,
\ee  
which is constructed from the energy scale of the harmonic repulsive force.
Accordingly, the viscosity is given in units of $\xi_{\rm n} /a$. Finally, the temperature is expressed in units of $\epsilon_{\rm n}/k_{\rm B}$.

We integrate the equations of motion with an Euler algorithm with an integration timestep $\Delta t = 0.1 t_0$. The accuracy of the numerical integration is confirmed by decreasing the time step at some selected state points. Because the system is highly overdamped and subject to thermal motion with white Gaussian noise, numerical errors do not accumulate over time and numerical integration is found to be quite stable. 

A typical simulation is decomposed into two parts. We first run the simulation at the chosen state point during some equilibration time, $t_{\rm eq}$. This is followed by the production run with a duration $t_{\rm sim}$.
We set $t_{\rm eq}=1/\gdot$, so that the system has effectively been deformed by 100\% before any measurement. The duration of the production run to analyse the data is set as $t_{\rm sim}=9/\gdot$ (for $N=1000$) and $t_{\rm sim}=1/\gdot$ (for $N=10000$).

We are interested in recording the flow curves describing the evolution of the viscosity $\eta$ with the imposed shear rate $\gdot$ for a given state point. To this end we measure the shear stress $\sigma_{xy}$ directly in the simulation as a time average 
$\sigma_{xy} = (1/t_{\rm sim}) \int_{t_{\rm eq}}^{t_{\rm eq}+t_{\rm sim}}{\rm d}t \hat{\sigma}_{xy} (t)$, where the instantaneous stress is given by the usual expression  $\hat{\sigma}_{xy}(t)=-\frac{1}{V(t)}\sum_{k<j}(x_j-x_k) 
(f_{yjk}^{\rm nor}+f_{yjk}^{\rm tan})$, where $f_{y jk}^{\rm nor}$ and $f_{y jk}^{\rm tan}$ are the $y$ components of the normal and tangential forces acting between particles $j$ and $k$. Once the shear stress is measured at an imposed shear rate, we directly deduce the viscosity $\eta = \sigma_{xy} / \gdot$. We typically record the flow curves by successively decreasing $\gdot$ from the highest to the lowest studied value using the final configuration of a given $\gdot$ as initial condition for the equilibration for the next $\gdot$ value. In case where we observe a discontinuous evolution of the flow curve, the studied discontinuous shear-thickening, we also run upward runs where we successively increase the shear rate using the same equilibration/production procedure as described above. As discussed below and found in earlier studies~\cite{otsuki11,heussinger}, these two procedures often result in hysteretic behaviour when discontinuous shear-thickening is present. 

We varied the packing fraction over a broad range, from the dilute fluid at $\phi=0.50$ up to jammed packings at $\phi=0.70$. The temperature was varied from large values, $T=10^{-5}$ down to very small ones, $T=10^{-10}$, and the athermal case $T=0$ was also studied independently, both with or without frictional forces. These additional data will be more extensively analysed in a future article~\cite{preliminary}. The friction coefficient is also varied over a broad range from $\mu_{\rm C} = 10^{-2}$ where packings are nearly frictionless, up to $\mu_{\rm C}=10$ which almost represents the limit of large frictional coefficients (few experimental systems can actually reach that limit). We mostly report numerical results for the intermediate value $\mu_{\rm C}=1.0$ since changing the value of the friction coefficient does not affect our results in any qualitative way.

\subsection{Important control parameters}

As described above, the studied model is described by only few control parameters. The reason is that we have decided to exclude several physical ingredients from our modelling. The motivation is obvious, as this reduces considerably the parameter space, and the physical behaviour of the model can be fully explored without the need to `fix' by hand several additional parameters. Note in particular that we do not introduce any additional repulsive force between the particles, mimicking for instance electrostatic interactions or polymeric degrees of freedom at the surface of colloids. This choice thus differs from most earlier numerical studies~\cite{seto13}. 

For the unsheared system, we simply need to fix the competition between the amplitude of thermal and elastic forces, for a given value of the friction coefficient. This defines a static `state point'. For the constant volume simulations, a state point is characterized by temperature and volume fraction, $(T, \phi)$. For the constant pressure simulations the state point is instead defined by temperature and pressure, $(T, P)$. Because temperature is expressed in units given by the energy scale of the harmonic repulsive force, $[T] = \epsilon_{\rm n} / k_{\rm B}$, the value of the temperature actually controls the effective softness of the particles, as the kinetic energy $k_{\rm B} T$ quantifies how much pairs of particles can overlap, the typical overlap being $\delta \sim a \sqrt{ k_B T / \epsilon_{\rm n}}$, which can be easily derived from the interaction force in Eq.~(\ref{force}).   

In previous work, the physics of shear-thickening was often studied using `hard' spheres, but the numerical integration scheme used in those papers was relying on approximating the hard sphere potential with a softer one allowing for a finite overlap between particles~\cite{seto13,mari14,mari15}. In most cases, the maximum allowed overlap was kept to a constant value independently of the external parameters. In our language, this corresponds to a particle softness that would change with the shear rate, and thus to an ill-defined pair potential. Whereas this approximation is presumably irrelevant for the athermal studies in Refs.~\cite{seto13,mari14}, this is not the case when thermal fluctuations are included~\cite{mari15}, since the introduction of an effective softness then directly competes in an unwanted way with Brownian forces. Therefore the quantitative conclusions from Ref.~\cite{mari15} need revision, since they do not apply to hard spheres, nor to any soft potential. In our work, the hard sphere limit is obtained by taking the limits $T \to 0$ and $\gdot \to 0$. We shall see that this is a highly singular limit where the various regimes found for a finite softness all shift to infinite $\gdot$, but with different scaling forms. The numerical value of the cutoff used in Ref.~\cite{mari15} corresponds in fact to the softest particles investigated here, indeed quite far from the hard sphere limit.  

Beyond the softness of the particles, the temperature also defines an important timescale~\cite{ikeda}, 
\be \tau_{\rm T} = \xi_{\rm n} a^2/ (k_{\rm B} T).
\ee 
This corresponds to the time it takes a single particle to diffuse over its own size $a$ using thermal fluctuations. This timescale $\tau_{\rm T}$, which is defined for the system at rest, becomes very relevant when the rheological behaviour is explored at finite shear rate, since one can then define the dimensionless P\'eclet number, 
\be 
\Pe = \gdot \tau_{\rm T}.
\ee 
This allows us to distinguish between the low-P\'eclet regime, $\Pe \ll 1$, where the shear rate is low enough that thermal motion may influence the dynamics, and the large P\'eclet regime, $\Pe \gg 1$, where thermal fluctuations are too slow to affect the physics and the system is essentially athermal. These two regimes are useful since they separate the thermal glass physics occurring at low P\'eclet from the athermal jamming physics taking place at large P\'eclet~\cite{ikeda,ikeda_soft}. The crossover between the two regimes occurs near $\Pe \approx  1$, which in our reduced numerical units corresponds to the simple expression $\gdot \approx T$. 

In previous work, the role of frictional forces was mainly investigated in the absence of thermal fluctuations, i.e. taking $\Pe = \infty$ from the outset. Here, we shall investigate how thermal forces may change that picture. A naive expectation is that the regime $\Pe <1$ is the only one affected by thermal forces, leaving the athermal regime $\Pe > 1$ essentially unaffected. We shall see that this is not correct, as thermal forces also introduce a typical force scale which can be estimated from the fluctuation-dissipation theorem and can be converted into an `instantaneous' thermal stress scale $\sigma^{\rm inst}_T$,
\be 
\sigma^{\rm inst}_T = \sigma_0  \sqrt{\frac{ k_{\rm B} T } {\epsilon_{\rm n}}}. 
\label{stress}
\ee 
Note that this instantaneous stress scale differs from the typical stress scale that should be used when discussing glassy solids for which the entropic contribution to the yield stress is important, which reads instead $\sigma_{\rm T} / \sigma_0 = k_{\rm B} T / \epsilon_{\rm n}$.
The thermal stress scale in Eq.~(\ref{stress}) does not appear in previous work discussing thermal effects near jamming~\cite{ikeda,mari15,ikeda_soft}.
We shall see that it plays a crucial role in the mechanism of shear-thickening in Brownian suspensions. Its mathematical expression makes it very clear that it is intimately linked to the particle softness. 
 
\section{Macroscopic flow curves}

\label{macroscopic}

\subsection{Density dependence of flow curves}

\begin{figure}
\psfig{file=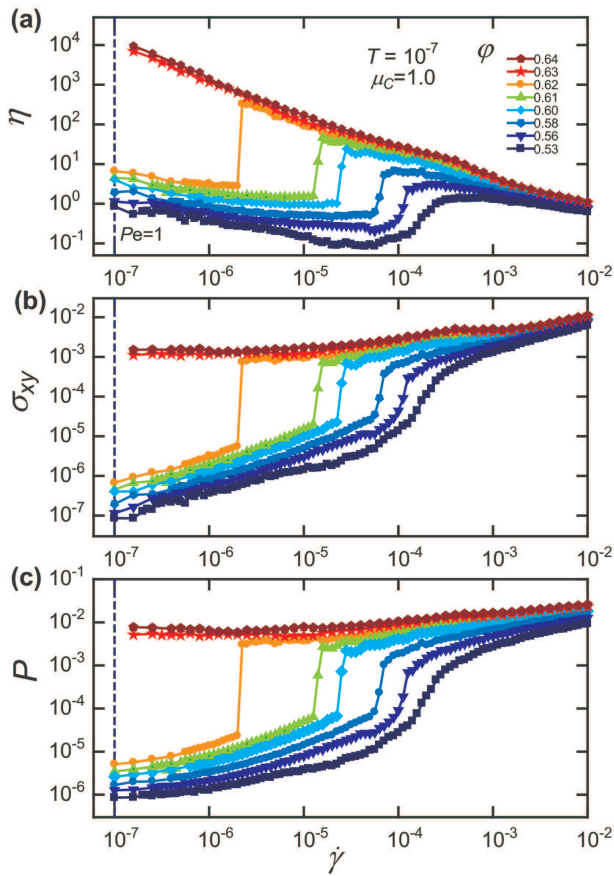,width=8.2cm}
\caption{\label{fig1} Behaviour of the system for $\mu_{\rm C}=1.0$ and $T=10^{-7}$, as the shear rate is deceased at constant packing fraction for $N=1000$ particles. 
(a) The flow curves $\eta(\gdot)$ display shear-thinning at small $\gdot$ followed by discontinuous shear-thickening at larger $\gdot$. 
(b) The same flow curves shown as $\sigma_{xy} (\gdot)$, where discontinuous shear-thickening now appears as a discontinuous stress jump. 
(c) The evolution of the pressure $P(\gdot)$ mirrors the one of the shear stress. In all panels, the vertical dashed line at $\gdot=10^{-7}$ locates the shear rate where $\Pe=1$.}
\end{figure}

We begin by showing typical flow curves obtained in presence of both a finite amount of thermal fluctuations, $T=10^{-7}$, and a finite friction coefficient, $\mu_{\rm C}=1.0$. We then vary the packing fraction from $\phi=0.53$ far below jamming, to $\phi=0.64$, very close to the jamming transition of frictionless particles (we have measured $\phi_J(\mu_{\rm C}=0)=0.647$). For this value of the friction coefficient, frictional particles jam near $\phi_J(\mu_{\rm C}=1.0) = 0.60$. For each packing fraction we vary the shear rate over a broad range, from $\gdot \sim10^{-7}$ to $\gdot = 10^{-2}$. We present our results in Fig.~\ref{fig1}, where we show the evolution of the viscosity, the shear stress, and the pressure as a function of $\gdot$. 

In each panel the limit between athermal and thermal regimes at $\Pe=1$ is indicated by a vertical dashed line at $\gdot = T = 10^{-7}$ so that most data are effectively taken in the athermal regime where $\Pe \gg 1$. In this athermal regime, the system would essentially display Newtonian rheology in the absence of frictional forces, as the system would be unjammed up until $\phi_J=0.647$. On the other hand, the athermal system with frictional forces would jam earlier and would display a finite yield stress above $\phi_J=0.60$.  

For each density we observe that the viscosity initially decreases with increasing $\gdot$. This shear-thinning regime has its origin in the thermal glassy physics of the system, and it corresponds to the shear-thinning observed in dense colloidal suspensions near glass transitions~\cite{ikeda}. 
As $\gdot$ is increased further towards the athermal regime, we observe for each density a strong and sharp shear-thickening occurring at a packing fraction dependent shear rate, from $\gdot \sim 10^{-4}$ at $\phi=0.53$ up to {$\gdot \sim 10^{-6}$ for $\phi=0.62$}. The thickening is large (about one order of magnitude) but continuous at $\phi=0.53$, and becomes even larger (about three orders of magnitude) and discontinuous for $\phi \ge 0.55$. 

Qualitatively, these flow curves are composed of the three different pieces of physics that the model incorporates: (i) shear-thinning at small $\gdot$ due to thermal rheology of dense suspensions, (ii) nearly Newtonian rheology at intermediate $\gdot$ as in athermal frictionless systems, (iii) yield stress rheology at even larger $\gdot$ as in athermal frictional systems. The transition between (i) and (ii) occurs near $\Pe = 1$, as expected~\cite{ikeda}, whereas the transition between (ii) and (iii) is very sharp and corresponds to the discontinuous shear-thickening that is the central topic of the present work. Strikingly, the transition between regimes (ii) and (iii), which is completely ruled by thermal fluctuations (as shown below) occurs deep in the athermal regime at $\Pe \gg 1$. This suggests that in the presence of frictional forces, the crossover between thermal and athermal rheologies near jamming becomes much more complicated than in the pure frictionless case~\cite{ikeda,ikeda_soft}.  

Overall, these flow curves are qualitatively similar to experimental observations in colloidal suspensions, with an interplay between shear-thinning at low $\gdot$ followed by a discontinuous shear-thickening behaviour at larger $\gdot$, these regimes being strongly dependent on the packing fraction. This shows that a simple model of soft repulsive particles with Brownian fluctuations and frictional forces is enough to reproduce this behaviour without the need to introduce any other interactions either of electrostatic or hydrodynamic origins. In the rest of the paper, we analyse this behaviour at the microscopic level, and provide quantitative measurements and predictions of how the various regimes depend on the external control parameters. 
 
In Fig.~\ref{fig1}b we show the same data in a different representation, $\sigma_{xy}$ versus $\gdot$, for the same conditions as in Fig.~\ref{fig1}a. This allows us to define two stress scales for the discontinuous shear-thickening as the stress jumps from one value to another as shear-thickening takes place, which we respectively name  
$\sigma_{\rm c}^{\rm L}$ and $\sigma_{\rm c}^{\rm H}$. Finally, we show the evolution of the pressure $P$ as a function of $\gdot$ in Fig.~\ref{fig1}c for the same set of parameters. As expected, the behaviour of the pressure closely follows the one of the shear stress, displaying in particular a discontinuous jump as the discontinuous shear-thickening transition is crossed. This upward jump of the pressure is the signature of dilatancy when using constant volume simulations. The system would like to dilate but the volume is kept constant by construction, and so the pressure increases instead. It is therefore clear that performing simulations that keep the pressure fixed must dramatically impact the behaviour of the system, as we confirm below in Sec.~\ref{pressure}.   

\subsection{Dynamic state diagram}

\begin{figure}
\psfig{file=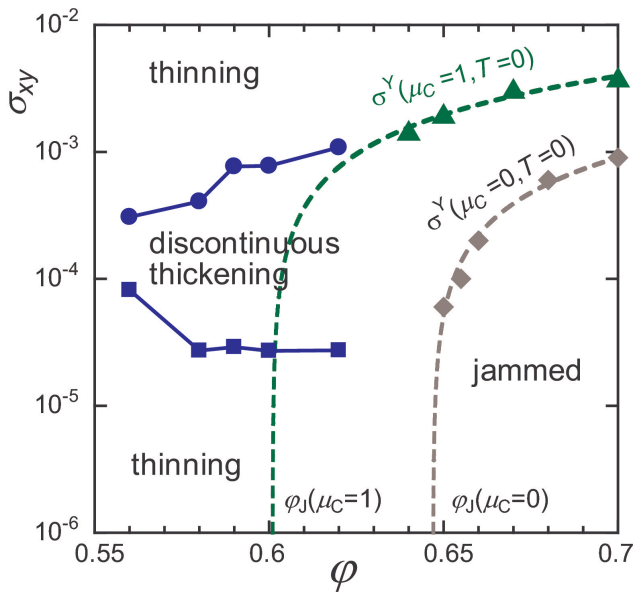,width=8.5cm}
\caption{ \label{fig2} Dynamic state diagram of multiple rheological regimes at $T=10^{-7}$ with $\mu_{\rm C}=1.0$ for $N=1000$. Below the jamming packing fraction of the frictionless case, a succession of thinning, discontinuous shear thickening, thinning rheology is observed as the stress increases. 
For reference we also show the yield stress evolution of both limits of athermal frictional and frictional soft spheres, fitted to a simple linear vanishing with the distance to jamming (dashed lines).} 
\end{figure}

We can summarize the various rheological regimes discussed in the flow curves in a dynamic state diagram, as this type of representation has often been used in previous experimental and numerical studies, e.g. Refs.~\cite{brown,seto13}. We present a stress-volume fraction phase diagram that allows us to incorporate the discontinuous shear-thickening physics into the broader context of the jamming transitions of frictional and frictionless systems. For the specific values $\mu_{\rm C}=1.0$ and $T=10^{-7}$ used in Fig.~\ref{fig1}, we report the boundaries $\sigma_{\rm c}^{\rm L}$ and $\sigma_{\rm c}^{\rm H}$ of the discontinuous stress jumps as a function of the volume fraction, see Fig.~\ref{fig2}. According to Fig.~\ref{fig1}, we find discontinuous shear-thickening in the region $0.55 \leq \phi \leq 0.63$ for $T=10^{-7}$. For densities lower than $\phi=0.55$, the system exhibits a continuous shear thickening, as found before.  

We can compare these stress scales to various important stress scales. In the athermal limit, the system may become jammed either with or without frictional forces. In both cases, a finite yield stress would emerge continuously at the jamming transition. We have performed additional simulations~\cite{preliminary} at $T=0$ to directly measure those yield stresses at various densities above $\phi_J=0.60$ (with friction) and above $\phi_J=0.647$ (without friction).
We obtain the yield stress value $\sigma^{\rm Y}$ for each density by fitting the flow curves $\sigma_{xy}(\gdot)$ to the Herschel-Bulkley expression~\cite{Larson_book}: $\sigma_{xy}=\sigma^{\rm Y}+ a \gdot^{b}$ where $a$ is an adjustable parameter and $b=0.35$~\cite{preliminary}. We confirm that the yield stress increases continuously with the distance to jamming as $\sigma^{\rm Y} \propto (\phi-\phi_{\rm J})$ both with and without friction. Above the frictionless jamming packing fraction, the emergence of yield stress in the unsheared system prevents the observation of shear-thickening, as also found experimentally~\cite{brown2010}.  

The dynamic state diagram in Fig.~\ref{fig2} is clearly reminiscent of the one found in several experiments reporting discontinuous shear-thickening rheology, where a succession of thinning-thickening-thinning regimes occurs at low volume fraction when the stress increases, whereas jamming rheology is observed instead at larger volume fraction~\cite{brown} which prevents the emergence of shear-thickening. This confirms once again the close similarity between the findings from our simple model and a large number of experimental reports on dense colloidal suspensions.  

\section{Transition to frictional rheology controlled by thermal fluctuations}

\label{microscopic}

In the previous section, Sec.~\ref{macroscopic}, we presented direct evidence from a specific set of control parameters that our model of thermal soft spheres with frictional forces reproduces all known phenomenology of colloidal suspensions with discontinuous shear thickening. In this section, we provide microscopic insight into the physical origin of the thickening behaviour, and explain how the physics quantitatively varies with the external control parameters, thus fully elucidating the physics and the relevant crossover scales that characterizes the present computer model. We will demonstrate that discontinuous shear-thickening corresponds to a sharp transition from athermal frictionless to athermal frictional rheology, but controlled by thermal fluctuations and occurring deep in the athermal regime of large P\'eclet numbers. 

\subsection{Direct evidence from flow curves}

\begin{figure}
\psfig{file=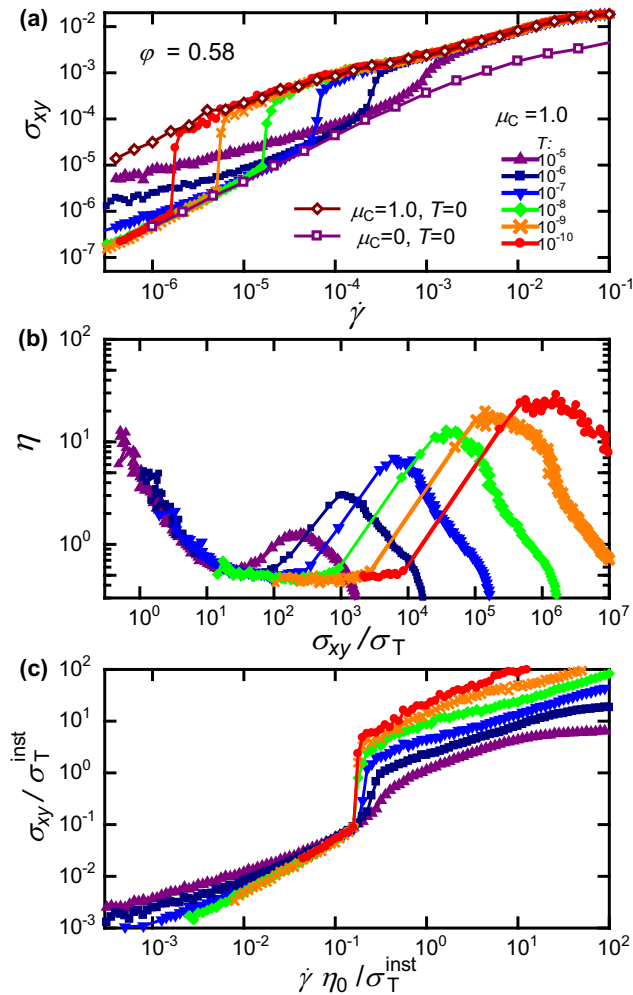,width=8.4cm}
\caption{\label{fig3} 
(a) Flow curves $\sigma_{xy}(\gdot)$ for $\varphi=0.58$ and $\mu_{\rm C}=1.0$ for several temperatures. The $T=0$ flow curves with $\mu_{\rm C}=1.0$ and $\mu_{\rm C}=0$ are highlighted with open symbols. 
(b) Same data shown using the representation $\eta$  as a function of the shear stress $\sigma_{xy}$ rescaled by the thermal stress $\sigma_{\rm T} = k_{\rm B}T/a^3$. This shows that the onset of shear-thickening occurs at a temperature dependent stress scale that is much larger than $k_{\rm B}T/a^3$.    
(c) Rescaling of the shear-thickening onset using the instantaneous thermal stress scale $\sigma_{\rm T}^{\rm inst}=\sigma_0\sqrt{k_{\rm B}T/\epsilon_{\rm n}}$, which shows that Brownian forces act as a repulsive force against frictional contacts.} 
\end{figure}

Our modelling of discontinuous shear thickening relies on the idea that all the ingredients within our model are needed to observe interesting thickening physics. To establish this quantitatively, we first show that the limit rheology obtained when temperature is strictly zero, both with and without frictional forces. 

In Fig.~{\ref{fig3}}a we compare the flow curves at $T=0$ for both frictionless ($\mu_{\rm C}=0$) and frictional ($\mu_{\rm C}=1.0$) cases for a packing fraction $\phi=0.58$. When $\gdot$ is small, the stress increases linearly with the shear rate, and Newtonian behaviour is obtained, with two distinct values of the viscosity. This is expected, as the packing fraction is below the jamming value for both cases. At larger shear rate, particle softness manifests itself, and weak shear-thinning rheology is obtained~\cite{teitel}. The important observation is that no shear-thickening is observed at any $\gdot$ when $T=0$. It is also known from earlier extensive work on thermal frictionless particles~\cite{ikeda,ikeda_soft} that discontinuous shear-thickening is not observed when $T>0$ and frictional forces are absent. Therefore, we conclude that thermal fluctuations and frictional forces need to act together to give rise to shear-thickening.

This is demonstrated in Fig.~{\ref{fig3}}a where we gradually increase the temperature from $T=0$ over a broad range, from $T=10^{-10}$ to $T=10^{-5}$, for a fixed value of the friction coefficient, $\mu_{\rm C}=1.0$. Although the variation of temperature is large (4 orders of magnitude), its overall amplitude remains weak, so that the particle softness varies over a range representative of hard sphere colloidal systems to softness such as the ones typically found in colloidal microgels.  For all values of the temperature, we now find that the viscosity follows the athermal frictionless rheology at small shear rates, and jumps discontinuously on the flow curve found for the athermal frictional limit. We observe deviations at small $\gdot$ from this ideal description for the two highest temperatures, see Fig.~\ref{fig3}a, but it is important to notice that they appear in the regime $\Pe < 1$, where indeed thermal fluctuations are expected to modify the rheology and make it shear-thinning due to the emergence of glassy physics. For all other state points, the rheology at finite $T$ closely follows the athermal limit rheology, frictionless or frictional. Thus, temperature simply serves in this regime as an activator for frictional forces but does not change the overall rheological picture.  

The clear conclusion from these data is that the rheology of thermal frictional particles exhibits a sharp crossover from the athermal frictionless to the athermal frictional rheology at a well-defined shear rate value. We find in addition that the jump towards frictional rheology occurs at a crossover shear rate that varies continuously with the amplitude of the temperature. This crossover shear rate varies by about 2-3 orders of magnitude as the temperature varies over 5 decades, which suggests that the crossover is not controlled by particle softness directly (proportional to $T$) but is rather controlled by the instantaneous thermal stress scale defined in Eq.~(\ref{stress}) (proportional to $\sqrt{T}$). 

In addition, we also notice that the crossover shear rate, for each flow curve, occurs very deep into the athermal regime at $\Pe \gg 1$. Our general conclusion is thus that the model exhibits a discontinuous shear-thickening rheology, at a crossover shear rate that lies deep within the athermal regime at large P\'eclet number, but the crossover itself is nevertheless governed by the intensity of thermal fluctuations. The shear-thickening itself is interpreted, as in previous models, as a sharp transition between frictionless and frictional rheologies, as confirmed below.

More quantitatively, we replot the data shown in Fig.~\ref{fig3}a using a representation where the evolution of the viscosity $\eta$ is shown as a function of the stress (as often done in shear-thickening studies), and we rescale the stress by the thermal stress scale $\sigma_{\rm T} = T$ (in reduced units). In this representation the shear-thinning regime at $\Pe<1$ becomes clearly visible at very low stress, $\sigma_{xy}/ \sigma_{\rm T} \sim 1$, as well as the transition to Newtonian athermal rheology at larger stress, $\sigma_{xy}/\sigma_{\rm T} \sim 10-100$. Interestingly, the onset stress scale for discontinuous shear-thickening occurs at a rescaled stress values $\sigma_{xy} / \sigma_{\rm T}$ that changes from about 50 for $T=10^{-5}$ to $10^4$ for $T=10^{-10}$. In the work of Mari {\it et al.} on `hard' spheres, the rescaled crossover stress scale for discontinuous shear-thickening due to thermal fluctuations is $\sigma_{xy} \approx 5 k_{\rm B} T /a^3$, which corresponds in our units to a softness regime that we have not simulated as it is too far from the physically relevant regime. This discrepancy presumably arises because Mari {\it et al.} vary the effective particle softness for each $\gdot$ to obtain their `hard sphere' limit. In fact, changing the softness corresponds to changing $k_{\rm B}T/\epsilon_{\rm n}$. Therefore, taking the envelope for the flow curves shown in Fig.~\ref{fig3}a might be semiquantitatively consistent to the results in Ref.~\cite{mari15}. When thermal fluctuations and particle softness are modelled in a consistent way, the conclusion by Mari {\it et al.} that another repulsive force is needed to account for experimental findings does not hold anymore. 

The direct proof that it is the instantaneous thermal stress $\sigma_{\rm T}^{\rm inst}$ that controls the shear-thickening crossover is offered in Fig.~\ref{fig3}c where we rescale the flow curves in panel (a) using $\sigma_{\rm T}^{\rm inst}$. It becomes clear that the onset of shear-thickening is indeed controlled by $\sigma_{\rm T}^{\rm inst}$. The physical interpretation is simple, as two spheres that are undergoing a frictional contact also experience a thermal force at each time step whose amplitude is given by $\sigma_{\rm T}^{\rm inst}$ and direction is random, see the equation of motion Eq.~(\ref{em}). As soon as this Brownian force overcomes the stress $\sigma_{xy}$ imposed by the external shear flow, then the frictional contact between the two spheres can get disrupted and the spheres effectively behave as if they were frictionless. This explanation agrees with the one in Ref.~\cite{mari15} that thermal fluctuations ``act as a repulsive force'', but it further quantifies it for the pair interaction studied in our work. If we restore momentarily our physical units, this means that the stress scale for the onset of discontinuous shear-thickening is given by 
\be
\sigma_{\rm c} \approx \sigma_{\rm T}^{\rm inst} = \sigma_0 \sqrt{k_B T / \epsilon_{\rm n}}.
\label{onsetstress}
\ee
This expression clearly demonstrates that the onset stress scale for discontinuous shear-thickening depends directly on the particle softness, with the hard sphere limit being a singular limit. Another conclusion is that $\sigma_c \gg \sigma_{\rm T}$, which explains why the shear-thickening is observed in the athermal rheological regime at large $\Pe$, but is nevertheless still directly controlled by Brownian forces.  

\subsection{Direct evidence from the microstructure}

\begin{figure}
\psfig{file=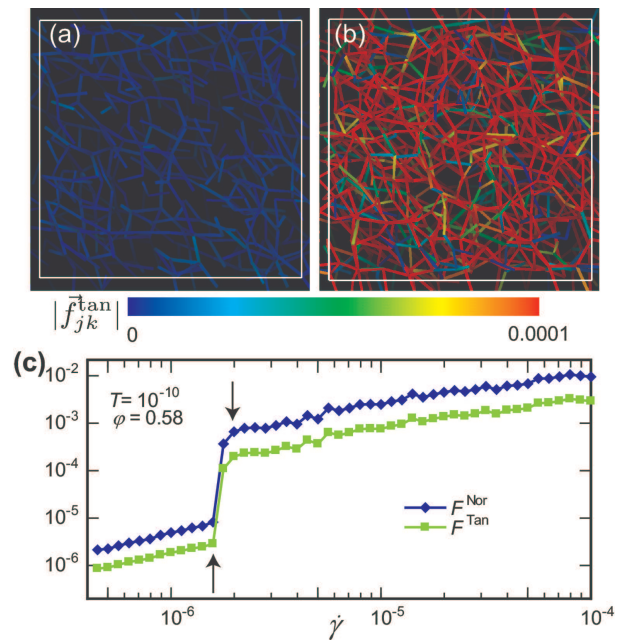,width=8cm}
\caption{\label{fig4}
Snapshots of tangential force contacts in constant volume simulation at $\phi = 0.58$ for $T=10^{-10}$ and $\mu_{\rm C}=1.0$ when (a) $\gdot = 2.00 \times 10^{-6} $  and (b) $\gdot = 1.58\times  10^{-6}$.
A bond between particles $j$ and $k$ is shown when $|\vec{f}_{jk}^{\rm tan}| >0$, and its color represents the  strength of $|\vec{f}_{jk}^{\rm tan}|$.
(c) The averaged frictional contact force 
$F^{\rm Tan} (=\sum_{jk}|\vec{f}_{jk}^{\rm tan}|/N)$ and normal contact force $F^{\rm Nor} (=\sum_{jk}|\vec{f}_{jk}^{\rm nor}|/N)$ with $\gdot$ for the same parameters, with arrows locating the snapshots in panels (a) and (b). The frictional contact forces have a discontinuous jump.}
\end{figure}

We now present direct evidence from the microstructure of the sheared packings that the discontinuous shear-thickening observed in the macroscopic flow curves corresponds, as in previous work, to a transition from the frictionless to the frictional rheology as the shear stress is increased. 
 
In Figs.~\ref{fig4}a,b we show a snapshot to depict the tangential force contacts obtained for $\phi = 0.58$ with $\mu_{\rm C}=1.0$ at $T=10^{-10}$ at two different shear rates, $\gdot = 1.58\times 10^{-6}$ in (a)  and  $\gdot = 2.00\times  10^{-6}$ in (b). A bond between two particles $j$ and $k$ is represented when the strength of the tangential force between them is non-zero, $|\vec{f}_{jk}^{\rm tan}|>0$. The snapshots are representative of the steady state behaviour for the given parameters. For $\gdot =  1.58 \times 10^{-6}$ shown in Fig. \ref{fig4}a, a small number of weak frictional contacts is observed. On the other hand, when the shear rate is increased by a small amount to $\gdot =  2.00 \times 10^{-6}$, a larger number of frictional contacts are now mobilised, and have a much larger strength. These visual impressions are reinforced by the quantitative measurement shown in Fig.~\ref{fig4}c, where we present the shear rate dependence of the averaged frictional contact forces for one particle, expressed as $F^{\rm Tan}=\sum_{jk} |\vec{f}_{jk}^{\rm tan}|/N$. 
A similar evolution is observed for the normal component of these forces $F^{\rm Nor}=\sum_{jk} |\vec{f}_{jk}^{\rm nor}|/N$.
A clear discontinuous jump is observed for $F^{\rm Tan}$ and $F^{\rm Nor}$, which mirror the discontinuous jump observed in the stress and in the pressure in Figs.~\ref{fig1}b,c.
The two arrows in Fig.~\ref{fig4}c correspond to the two snapshots shown in Figs.~\ref{fig4}a,b, on both sides of the viscosity discontinuity. 
From these measurements, we confirm that the microscopic origin of discontinuous shear thickening is a transition from frictionless to frictional flow regimes, since tangential forces get mobilised across the transition.

\subsection{Direct evidence from granular rheology}

\begin{figure}
\psfig{file=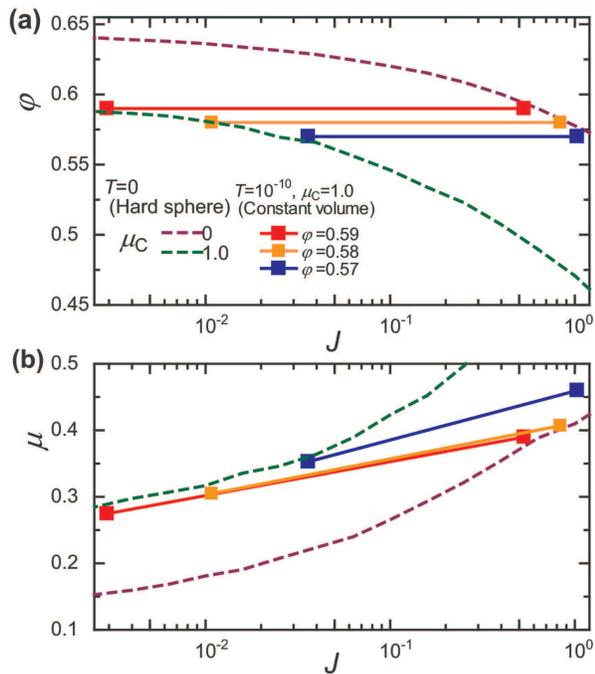,width=7.8cm}
\caption{\label{fig5}
Granular rheology representation of the discontinuous shear thickening.
The dashed lines represent 
(a) $\phi = \varphi(J)$ and (b) $\mu = \mu(J)$ for $N=1000$ in the hard sphere limit at $\mu_{\rm C}=0$ and $\mu_{\rm C}=1.0$ and $T=0$.
The symbols report the corresponding values on both sides of the discontinuous shear thickening transition measured at finite temperature $T=10^{-10}$ and various packing fractions for $\mu_{\rm C}=1.0$.}  
\end{figure}

It is useful to rephrase the above scenario of a frictionless-frictional transition to explain the discontinuous shear thickening in the language of hard granular materials. To this end, we introduce the viscous number~\cite{kawasaki14,boyer}
\begin{equation}
J = \frac{\gdot \eta_0}{P}.
\end{equation}
It is well established that in the limit of hard sphere interactions and in the absence of thermal fluctuations, the relevant dimensionless numbers are the packing fraction $\varphi$ and the (macroscopic) friction coefficient 
\begin{equation}
\mu=\frac{\sigma_{xy}}{P},
\end{equation}
and they become unique functions as the viscous number $J$. 
In Figs.~\ref{fig5}a,b, we plot the $\phi = \phi(J)$ and $\mu = \mu(J)$ curves obtained at $T=0$ for both $\mu_{\rm C}=1.0$ and $\mu_{\rm C}=0$. As in our previous work using athermal soft spheres, we have taken the hard sphere limit to measure the evolution of the granular rheology~\cite{kawasaki14}. As expected, the two rheologies have similar functional forms, but they differ quantitatively. In the $J \to 0$ limit, the packing fraction reaches different jamming densities ($\phi_{\rm J}(\mu_{\rm C}=1.0) \sim 0.60$ and  $\phi_{\rm J}(\mu_{\rm C} =0 ) \sim 0.647$), as already observed in the state diagram in Fig.~\ref{fig2} above. Similarly the friction coefficient $\mu_{\rm J} = \mu(J \to 0)$ differs in both cases ($\mu_{\rm J}(\mu_{\rm C}=1.0) \sim 0.23$ and  $\mu_{\rm J}(\mu_{\rm C} = 0 ) \sim 0.11$). These values are consistent with earlier numerical determinations~\cite{kawasaki15,silbert02}. 

Having measured the granular rheology (in the athermal hard sphere limit), we now superimpose to those data the measured boundaries for the discontinuous shear-thickening observed in our constant volume simulations at finite $T$. For each volume fraction, we measure the values of the shear rate, shear stress and pressure at the viscosity discontinuity, and report those jumps in the $\phi(J)$ and $\mu(J)$ representation appropriate to hard grains. It is very clear from this representation that the discontinuous shear thickening represents a transition between the athermal rheologies of frictionless and frictional grains, which is only observed when thermal fluctuations are present. This results generalise to different packing fraction the similar observation reported in Fig.~\ref{fig1} for $\phi=0.58$.  

\section{Quantitative consequences for experiments}

\label{quantitative}

\subsection{Sketch of the flow curves}

In this subsection we summarize the above discussions on the nature of the discontinuous shear-thickening in our model of soft repulsive spheres in the presence of frictional forces and thermal fluctuations. Such modeling is useful to describe colloidal particles. The goal of this section is thus to recap the various flow regimes expected for such dense suspensions and how they evolve with external control parameters, in order to be able to make contact with experimental studies in the following subsection.

\begin{figure}
\psfig{file=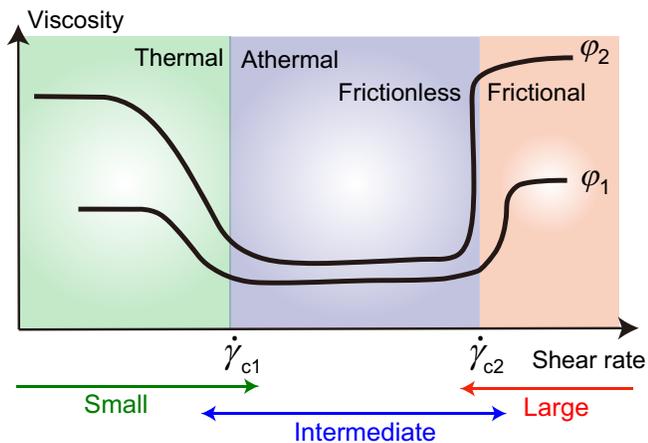,width=8.5cm}
\caption{\label{fig6} Sketch of the flow curves for dense suspensions of Brownian colloidal particles with frictional forces. This comprises a thermal-athermal crossover at $\gdot_{c1}$, and a frictionless-frictional crossover at $\gdot_{c2}$, as given in Eqs.~(\ref{gdot1}, \ref{gdot2}).
Glassy colloidal rheology is observed for $\gdot < \gdot_{c1}$,
athermal frictionless rheology for $\gdot_{c1} < \gdot < \gdot_{c2}$, and athermal frictional rheology for $\gdot_{c2} < \gdot$. Small, intermediate and large particles allow experimentalists to cover the various regimes exhibited by our model.}
\end{figure}

The model under study possesses two main crossovers: (i) from thermal to athermal rheology, which occurs for a P\'eclet number of order unity, 
\begin{equation}
\gdot_{c1} \approx 1/\tau_{T} = \frac{k_{\rm B}T}{\eta_0 a^3} =  \frac{\sigma_0}{\eta_0} {\frac{k_{\rm B}T}{\epsilon_{\rm n}}} ,
\label{gdot1}
\end{equation}
and (ii) from frictionless to frictional rheology, which occurs for shear rates scaling as 
\begin{equation}
\gdot_{c2} \approx \frac{\sigma_c}{\eta_0} = \frac{\sigma_0}{\eta_0} \sqrt{\frac{k_{\rm B}T}{\epsilon_{\rm n}}}. 
\label{gdot2}
\end{equation}
In the above two expressions, we neglected sub-leading density dependences easily obtained by matching more precisely the various regimes, to illustrate the leading dependence upon the control parameters of these two crossovers. Since typical particle softness in colloidal experiments correspond to quite small reduced temperatures~\cite{ikeda_soft}, the above expressions imply $\gdot_{c1} \ll \gdot_{c2}$, so that the succession of rheological regimes as the shear rate is ramped up is  
(1) thermal frictionless, (2) athermal frictionless, (3) athermal frictional.
The behaviour of the shear viscosity for dense suspensions across these three regimes as well as its evolution with the packing fraction is sketched in Fig.~\ref{fig6}. Once the various regimes (1), (2), and (3) are properly ordered, these flow curves simply concatenate previously known rheological behaviours for these three physical situations. 

In the first regime (1), the viscosity is Newtonian at very small shear rate, with a value controlled by the approach to the colloidal glass transition. As the shear rate increases, a shear-thinning is observed when the timescale set by the shear rate competes with the (slow) structural relaxation time of the glassy colloidal suspension~\cite{yamamoto98}, usually called ``alpha-relaxation time'' and noted $\tau_\alpha$. This shear-thinning regime is then observed up until $\gdot$ reaches the thermal-athermal crossover, $\gdot \sim \gdot_{c1}$.   

For $\gdot_{c1} < \gdot < \gdot_{c2}$, the system enters the athermal frictionless regime (2). In this regime the rheology is Newtonian again~\cite{kawasaki15}, and the viscosity is uniquely controlled by the packing fraction, but its value is now set by the distance to the frictionless jamming transition that occurs at $\phi_J(\mu_{\rm C} = 0)$. This Newtonian regime is sketched in Fig.~\ref{fig6}.

For soft particles, increasing the shear rate would drive the system to a shear-thinning regime~\cite{kawasaki14} whereas for hard particles the Newtonian regime would extend to arbitrarily large shear rates. Instead, due to the presence of frictional forces, a sharp shear-thickening transition is observed at $\gdot \approx \gdot_{c2}$ which drives the system into the athermal frictional regime (3), so that for $\gdot > \gdot_{c2}$ the rheology is again Newtonian~\cite{silbert10,boyer}, but now with a shear viscosity that is controlled by the distance to the frictional jamming density $\phi_J(\mu_{\rm C}=1.0)$. The inequality  $\phi_J(\mu_{\rm C}=1.0) < \phi_J(\mu_{\rm C}=0)$ is the reason why the viscosity increases sharply across $\gdot_{c2}$. This final Newtonian regime, accessed through a sharp increase of the viscosity, is also sketched in Fig.~\ref{fig6}.  

For modestly soft colloidal particles (typical values are given below), the crossovers between the various regimes sketched in Fig.~\ref{fig6} may cover several orders of magnitude, and maybe difficult to observe in a single experiment. As suggested already for the thermal-athermal crossover \cite{ikeda}, and realised already in recent experiments \cite{poon}, it may be better to use a range of materials with varying particle sizes to cover the various regimes depicted in Fig.~\ref{fig6}. Since the microscopic timescale $\tau_0$ that sets the shear rate becomes larger for larger particles, experiments performed with large particles will typically only access ``large'' shear rates (i.e. large $\gdot \tau_0$), whereas small particles by construction explore ``small'' shear rates (i.e. small $\gdot \tau_0$) in the sketch of Fig.~\ref{fig6}.

Thus, experiments using small colloidal particles (e.g. $\approx 100$-$200{\rm nm}$) will typically only cover the thermal regime, as most ``hard sphere'' colloidal experiments indeed do~\cite{petekidis}. Slightly larger particles ($\approx 200$nm-1$\mu{\rm m}$) would be needed to conveniently probe the thermal-athermal crossover, and particles from a sub-micron up to a few microns would be needed to observe the shear-thickening regime. Finally, extremely large particles would only explore to the athermal frictional regime, and no shear-thickening transition would be observed in this purely granular regime because it would take place at immeasurably small shear rates.

Experimentalists are not short of additional ideas to explore the various regimes shown in Fig.~\ref{fig6}. For instance, one can use large granular particles, and introduce an additional `repulsive' force between the particles, that would play the same as thermal fluctuations to perform shear-thickening experiments with large grains. In this approach, this amounts to shifting the crossover $\gdot_{c2}$ back inside the experimental window. Another way to play with timescales is to use solvents with varying shear viscosities, since $\eta_0$ also enters the crossovers in Eqs.~(\ref{gdot1}, \ref{gdot2}).  

\begin{table*}
\begin{tabular}{| c | ccccc |} \hline
 &  Latex emulsion \cite{wagner_physicstoday}  & Silica colloid \cite{WagnerJCP} & PMMA colloid (small)  \cite{poon} & PMMA colloid (large) \cite{poon}&  Quartz particle\cite{fernandez}   \\ \hline
$a$[m] & $2.5\times 10^{-7}$  & $6.06\times 10^{-7}$& $4.04\times 10^{-7}$& $3.77\times 10^{-6}$ &	$1.2\times 10^{-5}$  \\  
$\eta_{\rm s} $[Pa s] & $1\times 10^{-3}$  & $5.44\times10^{-3}$& $2.4 \times  10^{-4}$	& {$2.8 \times  10^{-4}$}& $1\times 10^{-3}$  \\
$\sigma_{\rm c}^{\rm Exp} $ [Pa]  & {$5\times 10^2$} & 50 & $2\times 10^2$	&1.5  &1.6	\\ \hline
$k_{\rm B}T/\epsilon_{\rm n} $ & {$2.8\times 10^{-7}$}  &$1.4\times10^{-7}$& $9.9\times10^{-8}$ &$ 2.7 \times 10^{-9}$& $2.2\times10^{-12}$   \\ 
\hline 
\end{tabular}
\caption{\label{tb:table} Sets of experimental parameters: particle sizes $a$, solvent viscosity $\eta_{\rm s}$ , onset stress of shear thickening  $\sigma_{\rm c}^{\rm Exp}$ for 
latex emulsion~\cite{wagner_physicstoday}, silica colloid~\cite{WagnerJCP}, PMMA colloids with different sizes~\cite{poon}, and quartz particle~\cite{fernandez}.  
The corresponding dimensionless temperatures (i.e. particle softness) $k_{\rm B}T/\epsilon_{\rm n}$ obtained from Eq.~(\ref{temp}) are shown. }
\end{table*}

A final, but important, point we want to make about Fig.~\ref{fig6} concerns the hard sphere limit of the soft repulsive potential we use. There is no doubt that hard spheres play a large role in both liquid state theory for simple fluids, but also in rheological studies for both colloidal materials and computer simulations. This is of course due to the simple functional form of the potential. An obvious drawback is that the pair potential changes abruptly from 0 to $\infty$ at the particle diameter, i.e. there is no intrinsic lengthscale to smooth out the effect of the repulsive, which of course cannot true in a real colloidal suspension. Still, it is interesting to analyse the effect of the hard sphere limit in the sketch of Fig.~\ref{fig6}. For colloidal hard spheres where thermal fluctuations remain relevant (i.e. small to intermediate hard sphere colloids), experiments would be able to probe the thermal regime (for small colloids~\cite{petekidis}) and the thermal-athermal crossover for somewhat larger colloids~\cite{poon}. However, recalling the expression of $\gdot_{c1}$ and $\gdot_{c2}$ we realise that as the rescaled temperature goes to zero (and particles behave as hard spheres), these two shear rates become infinitely far from each other. Thus, for true Brownian hard spheres, it is impossible to observe $\gdot_{c1}$ and $\gdot_{c2}$ in a single experiment. The reason is quite simple (but as mentioned above totally unphysical!): frictional contacts between true hard spheres can not survive in the presence of an infinitesimal Brownian force, so that as soon as $T>0$ shear-thickening in Brownian hard spheres only occurs at infinitely large shear rates. Therefore, strictly speaking Brownian hard spheres cannot undergo shear-thickening and a finite amount of particle softness is needed to trigger this effect. Another way to trigger shear-thickening for hard spheres is of course to introduce another repulsive force that is not of thermal origin, such as van der Waals interactions~\cite{poon}, or an artificial repulsive force~\cite{mari14}. Qualitatively, these can also be seen as endowing hard spheres with a finite softness, or a finite surface rugosity. A corollary is that Brownian hard spheres with frictional forces represent such a singular limit that they cannot be used to understand discontinuous shear-thickening.

\subsection{Quantitative comparison to experiments}

In this subsection, we compare the flow curves obtained in our simulations to several experiments realised with model systems in order to assess the quantitative validity of the numerical model.
We have analysed more carefully experimental data in latex emulsion~\cite{wagner_physicstoday}, silica colloid~\cite{WagnerJCP}, nearly hard sphere PMMA colloids with multiple particle sizes~\cite{poon}, and quartz particle~\cite{fernandez}. 
In our model, the dimensionless temperature $k_{\rm B}T/\epsilon_{\rm n}$ is the central control parameter to determine the shape of the flow curve and the various crossovers. The value of the friction coefficient also plays a role, of course, but it mainly affects the values of the packing fraction where discontinuous shear-thickening is prominent. Given the large uncertainty about packing fraction determination in colloidal experiments, keeping the friction coefficient fixed is reasonable.  

In a first step, we determine for each experiment the optimal value of $k_{\rm B}T/\epsilon_{\rm n}$ that allows our model to best reproduce the full range of experimental flow curves. In our model, the onset stress of the shear thickening is estimated as 
$\sigma_{\rm c} \sim \sigma_{\rm T}^{\rm inst} =  (\epsilon_{\rm n}/a^3) \sqrt{ k_{\rm B}T/\epsilon_{\rm n}  }$, see Eq.~(\ref{onsetstress}).
In experiments, this onset stress (called $\sigma_{\rm c}^{\rm Exp}$) can be measured directly and can be used as a input value for our comparisons.
By using it, we can estimate the dimensionless temperature as
\be
\frac{k_{\rm B}T}{\epsilon_{\rm n}} = \left( \frac{k_{\rm B}T}{  {\sigma_c^{\rm Exp}} a^3} \right)^2,
\label{temp}
\ee
where $k_{\rm B} = 1.38065 \times 10^{-23} {\rm m}^2 \ {\rm kg}\ {\rm s}^{-2}\ {\rm K}^{-1}$
and $T=300 {\rm K}$ (room temperature). 
The values of $\sigma_c^{\rm Exp}$ and the diameter of the particles $a$ are shown in Table \ref{tb:table}. They are then used in Eq.~(\ref{temp}) to provide an estimate of the dimensionless temperature $k_{\rm B}T/\epsilon_{\rm n}$ for all experiments, as shown in Table \ref{tb:table}.

Unsurprisingly we find that the estimated dimensionless temperature is strongly correlated with the particle size $a$~\cite{ikeda_soft}. For instance, for latex particles whose diameter is rather small ($a=250 {\rm nm}$), 
$k_{\rm B}T/\epsilon_{\rm n} \sim 10^{-7}$, whereas for the quartz particles whose diameter is quite large ($a=12 \mu{\rm m}$), the dimensionless temperature becomes small, $k_{\rm B}T/\epsilon_{\rm n} \sim 10^{-12}$, since indeed the thermal regime ($\Pe <1$) is out of the experimental observation window
(as is often the case with large granular particles).
As reported in Ref.~\cite{poon}, we also confirm that $\sigma_c^{\rm Exp} \propto a^{-2}$ is well obeyed in experimental work. Using Eq.~(\ref{temp}), we can estimate the relation between the energy scale $\epsilon_{\rm n}$ and particle size $a$, which reads 
\be
\epsilon_{\rm n} =  \frac{({\sigma_c^{\rm Exp}} a^3)^2}{k_{\rm B}T} \propto a^{2}.
\label{stiff}
\ee 
This relation between particle softness and particle size was in fact already found via another physical argument in Ref.~\cite{ikeda_soft}, where the scaling behaviour of the athermal yield stress value is discussed in the context of glass and jamming transitions.  
In fact, the relation $\epsilon_{\rm n} \propto a^{2}$ can be explained, physically, via a reasoning analogous to the Laplace pressure derivation~\cite{ikeda_soft}, as detailed below in Appendix \ref{ref:laplace}.

\begin{figure*}
\psfig{file=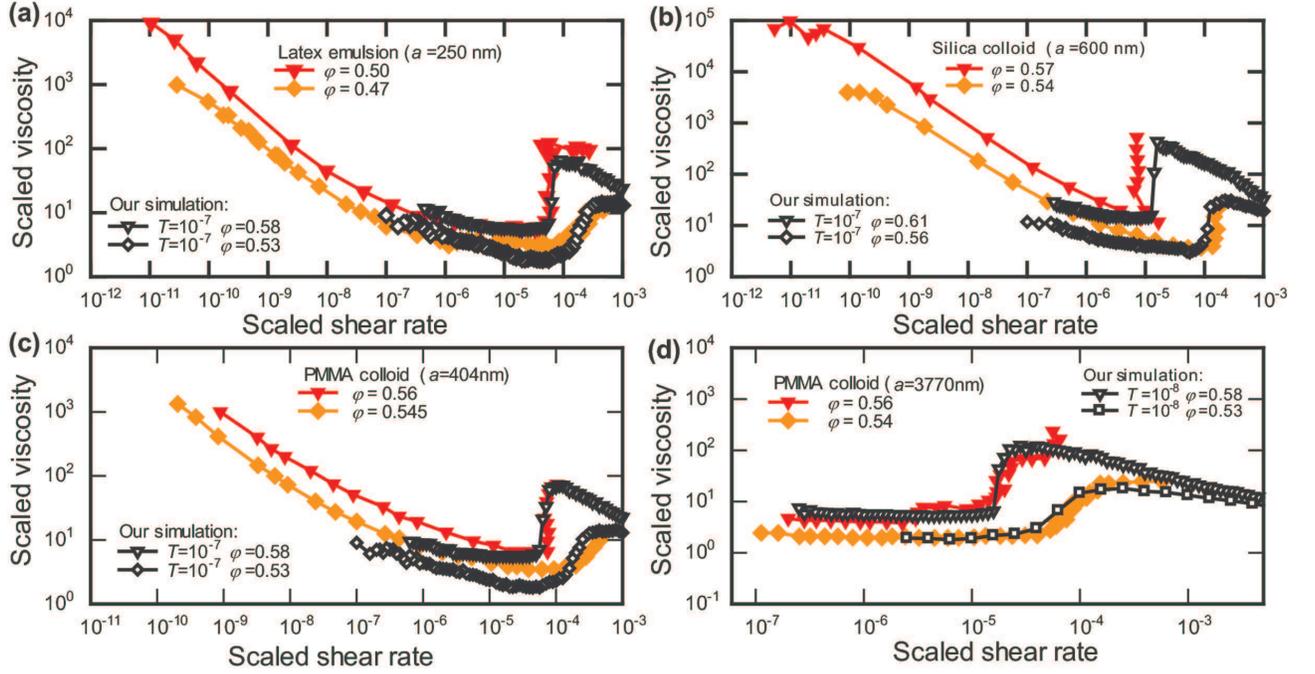,width=17.cm}
\caption{\label{experiment} Quantitative comparison of the flow curves obtained from experiments with our simulations.
(a) Latex dispersions~\cite{wagner_physicstoday} ($a=250$nm) for $\phi= 0.47$ and 0.50, compared with numerical flow curves for $\phi= 0.58$ and 0.53 at $T =10^{-7}$.  
(b) Small colloidal dispersions~\cite{WagnerJCP} ($a=600$nm) for $\phi= 0.57$ and 0.54, compared with numerical flow curves for $\phi= 0.61$ and 0.56
at $T =10^{-7}$.   
(c) Small colloidal dispersions~\cite{poon} ($a=404$nm) for $\phi= 0.56$ and 0.545, compared with numerical flow curves for $\phi= 0.58$ and 0.53
at $T =10^{-7}$.  
(d) Large colloidal dispersions~\cite{poon} ($a=3770$nm) for $\phi= 0.56$ and 0.54, compared with numerical flow curves for $\phi= 0.58$ and 0.53
at $T =10^{-8}$. 
Viscosities are scaled using $\eta/\eta_{\rm s}^{\rm eff}$ in experiments where $\eta_{\rm s}^{\rm eff} = 10 \eta_{\rm s}$ is an effective solvent viscosity. For simulations, we show $(\eta+\eta_{\rm s})/\eta_{\rm s}$, where $\eta_{s} = \eta_0 = \xi_{\rm n}/(3\pi a)$. 
The shear rates are scaled as $\gdot \tau_{\rm s}$ where $\tau_{\rm s} = 3\pi \eta_{\rm s}^{\rm eff} a^2/\epsilon_{\rm n}$ in experiments and $\tau_{\rm s} = t_0 = 3\pi \eta_{\rm s} a^3/\epsilon_{\rm n}$ in simulations.}
\label{sketch}
\end{figure*}

Armed with estimates of the dimensionless temperature relevant to describe each experimental system, we can directly compare the family of flow curves obtained in experiments and simulations. In Fig.~\ref{experiment}, we show experimental flow curves of latex~\cite{wagner_physicstoday}, silica colloids~\cite{WagnerJCP}, and multiple sizes of PMMA colloids~\cite{poon} dispersions. To compare experiments and simulations it is useful to scale the viscosity and the shear rate with the variables associated with the solvent viscosity $\eta_{\rm s}$, as given in Table~\ref{tb:table}. 

To complete the comparison between experiments and simulations there are two adjustements that are needed to get fully quantitative agreement. 
The first adjustment is about the simulations where the `solvent' is actually altogether absent, and its effect is only felt through the viscous damping. 
A consequence is that the viscosity of the simulated system would vanish at low density. To correct for this effect, we empirically correct the measured viscosity for the simulated system by plotting instead the quantity $(\eta+\eta_{\rm s})/\eta_{\rm s}$ where $\eta_{\rm s} = \xi_{\rm n}/a$. As a result, the viscosity is unaffected when $\eta \gg \eta_{\rm s}$, but this quantity goes to 1 (not 0) at low density when $\eta \ll \eta_{\rm s}$. This empirical rescaling of course affects none of the scaling behaviour discussed above. For the simulations, the shear rate is simply rescaled as $\gdot t_0$ where $t_0 = \xi_{\rm n} a^2/\epsilon_{\rm n}$, as defined in Sec.~\ref{unit}. The second adjustement we need to make is also a quantitative one. We find that we need to introduce an effective solvent viscosity $\eta_{\rm s}^{\rm eff}$ for experiments to obtain perfect quantitative consistency with the simulations. In Fig.~\ref{experiment}, the experimental viscosity and shear rate are scaled as $\eta/\eta_{\rm s}^{\rm eff}$ and $\gdot \tau_{\rm s}$ respectively, where 
$\tau_{\rm s} = 3\pi \eta_{\rm s}^{\rm eff} a^3/\epsilon_{\rm n}$. We find that imposing 
$\eta_{\rm s}^{\rm eff} = 10 \eta_{\rm s}$ yields perfect agreement between experiments and simulations. The factor 10 we find suggests that the viscosity of the experimental suspensions is about 10 times larger than the corresponding one in the simulations, but this constant factor does not depend on the state point. This is thus only a prefactor, which demonstrates that the scaling behaviour is the same in experiments and simulations, apart from a numerical adjustment related to solvent physics, which presumably adds an additive hydrodynamic contribution to the measured experimental values. 

When these two minimal adjustments are done, we obtain the results shown in Fig.~\ref{experiment}, where we superimpose flow curves obtained in experiments in 4 different materials, and the ones obtained in our simple numerical model. Our central conclusion is that the full range of the experimental flow curves which contain several non-trivial flow regimes (thinning, Newtonian, thickening) are quantitatively reproduced by our numerical model. 

Notice finally that the numerical flow curves in Fig.~\ref{sketch} display shear-thinning behaviour at extremely  large shear rates, in the regime which is athermal and frictional. This is because we use a finite particle softness in the present model, whereas perfect hard spheres would instead display Newtonian behaviour in that regime, as commonly observed in large granular particles~\cite{boyer,hinch}.    

\section{Constant pressure simulations}

\label{pressure}

\subsection{Absence of shear-thickening when pressure is constant}

Using the representation adopted in experiments concerning hard granular matter in Fig.~\ref{fig5}, we realised that the discontinuous shear-thickening behaviour observed in constant volume simulations also corresponds to a discontinuous increase of the pressure as the shear rate is increased. This observation has two consequences that we wish to explore in this section. First, it seems to suggest that keeping the pressure fixed should forbid the shear-thickening behaviour. Second, there exists pressure values that are never explored in the course of constant volume simulations, as the pressure jumps discontinuously. One may thus wonder what would happen if a forbidden pressure value was applied to the system.    

\begin{figure}
\psfig{file=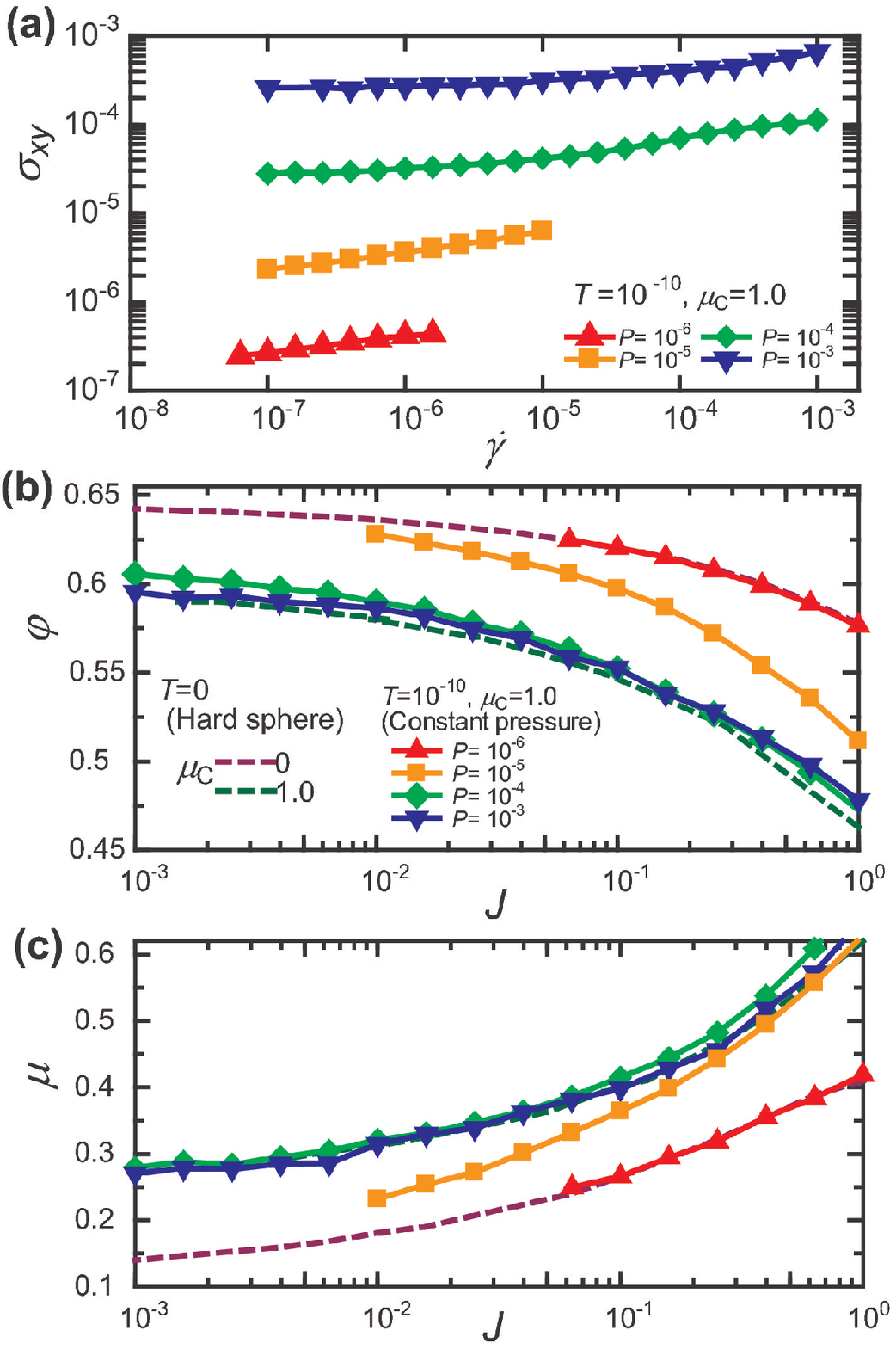,width=8.2cm}
\caption{\label{fig7}
(a) Flow curves $\sigma_{xy} = \sigma_{xy}(\gdot)$ measured during constant pressure simulations at various pressures for $N=1000$, $T=10^{-10}$, and $\mu_{\rm C}=1.0$.
(b) Same data replotted as $\varphi = \varphi(J)$ (b) and $\mu = \mu(J)$ (c) for the same parameters. The dashed lines represent the athermal hard sphere limit with and without friction, as in Fig.~\ref{fig5}. No discontinuous shear-thickening is observed in constant pressure simulations.} 
\end{figure}

We have performed simulations of our model for $T=10^{-10}$ and $\mu_{\rm C}=1.0$ using a constant pressure setup, as described in Sec.~\ref{constantP}. In that case, we vary the applied shear rate at constant $P$, and measure the shear stress to obtain the flow curves depicted in Fig.~\ref{fig7}a. For each pressure value, $P=10^{-3}, \cdots, 10^{-6}$, we observe a smooth increase of the shear stress with the shear rate, with no obvious discontinuity. Thus, it is immediately clear that discontinuous shear-thickening is fully prevented by in a constant pressure setup.  

We confirm this conclusion using the $\phi(J)$ and $\mu(J)$ representation of the granular rheology, see Figs.~\ref{fig7}b,c. Here again, we find that all data points evolve smoothly and do not show any discontinuous feature. Comparing the data obtained at constant pressure to the limiting case of athermal frictional and frictionless rheologies shown with dashed line demonstrates that very low pressure data tend to superimpose onto the frictionless rheology, whereas large pressure ones superimpose onto the frictional rheology. This effect simply confirms the physical picture of the discontinuous shear-thickening in terms of mobilisation at large enough shear stress of frictional forces. Imposing a large confining pressure ($P=10^{-3}, 10^{-4}$ in Fig.~\ref{fig7}) indeed mobilises the friction and the system appears frictional independently of the value of the shear rate, whereas for low pressures ($P=10^{-6}$ in Fig.~\ref{fig7}). As a result, the constant pressure system can only lie on one side of the thickening transition, but cannot cross it.
 
An interesting exception is the `forbidden' pressure value $P=10^{-5}$ in Fig.~\ref{fig7}.  According to the constant volume simulations for the same parameters, the system displays a pressure jump when discontinuous shear-thickening is observed and the value $P=10^{-5}$ is thus never reached. Hence we may wonder whether the system is somehow `unstable' is this pressure value is applied. 
However, the flow curve in Fig.~\ref{fig7}a shows that nothing really spectacular happens. Careful inspection of the simulations shows that there is no more temporal or spatial fluctuations for this pressure than for others. 
We conclude that the constant pressure setup is actually stable and does not give rise to any specific flow instability. When replotted in Figs.~\ref{fig7}b,c these data suggest a smooth crossover between frictionless rheology at small $J$ to frictional rheology at large $J$. 

\subsection{Do `S-shaped' flow curves exist?}

\begin{figure}
\psfig{file=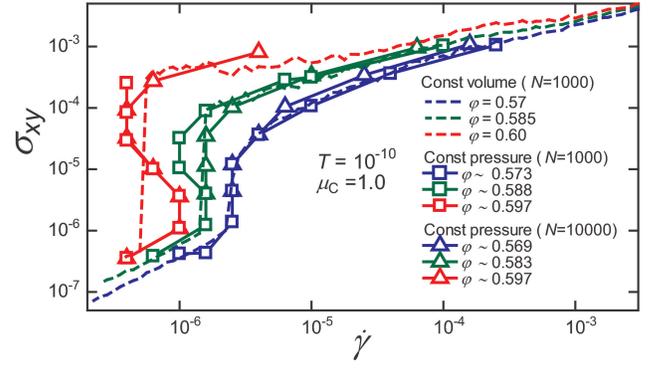,width=8.5cm}
\caption{\label{fig8}
The flow curves $\sigma_{xy} = \sigma_{xy}(\gdot)$ obtained via constant volume simulations for various $\varphi$ for several $\phi$ are shown with dashed lines. The flow curves obtained from the pressure controlled simulations with $N=1000$ (squares) and $10000$ (triangles) are also shown. Identical colors are used for identical volume fractions. All the data are obtained for $\mu_{\rm C}=1.0$ and $T=10^{-10}$. The discontinuous shear-thickening observed in constant volume simulations becomes and S-shaped flow curve for a finite $N$ simulation, that becomes more and more vertical as $N$ increases.}
\end{figure}

A surprising feature of the previous subsection is the observation that a pressure value that is unaccessible during the course of a constant volume simulation is instead easily accessed in a constant pressure setup, and provides a measurable state point where the viscosity and the volume fraction can be measured in steady state conditions. As a consequence, this setup allows us to `fill' the discontinuous viscosity jump observed at constant volume with additional data points. The exercise is quite tedious: to reconstruct a constant volume fraction flow curve, we need to very precisely adjust the pressure until the steady state value of $\phi$ is the one we wish to reach (we allow for a variation of $\phi$ of about $\pm 0.003$ to speed up the convergence of this iterative process). Once this agreement is achieved, we measure the viscosity. Repeating this analysis for a range of pressures, we can finally reconstruct the flow curves shown in Fig.~\ref{fig8}, which are obtained for $T=10^{-10}$ and $\mu_{\rm C}=1.0$. 

The first observation is that away from the shear-thickening discontinuity, constant pressure and constant volume simulations agree with one another, as expected. A more interesting observation is that for each density where a discontinuous shear-thickening was observed, we can now report the value of the viscosity in the middle of the sharp discontinuity. 
Strikingly, for a system size $N=1000$, we find that the discontinuity now becomes an `S-shaped' flow curve, which is obviously reminiscent of the van der Waals loop across a first-order phase transition in thermal equilibrium. 

A well-known feature of first-order phase transition is that the equilibrium loop becomes less pronounced as the system size increases. To test this intuition, we have repeated the constant pressure simulations described above for a much larger system with $N=10000$. The data are reported in Fig.~\ref{fig8} for the same parameters as before. Here, we observe that for larger systems, the S-shaped flow curve becomes much less pronounced than for the smaller system. Compare for instance the constant pressure simulations data at $\phi \sim 0.597$ for $N=1000$ and $N=10000$ in Fig.~\ref{fig8}. Thus, we expect that for an even larger system, the S-shaped curve would turn into an `I-shaped' flow curve, i.e. the non-monotonic behaviour would disappear altogether and constant pressure and constant volume simulations would finally agree in the thermodynamic limit. The disappearance of S-shaped curves with system size is consistent with the numerical analysis performed in Ref.~\cite{heussinger15}, which has been obtained in the context of a distinct microscopic model in two dimensions.

As an important remark on the issue of non-monotonic flow curves, we point out that the seemingly unstable flow curves displayed in Fig.~\ref{fig8} are actually reconstructed one point at a time from flow curves such as the ones reported in Fig.~\ref{fig7}a, which are instead totally featureless and completely stable. Thus, there is no fundamental reason for the flow curves in Fig.~\ref{fig8} to be unstable, and each point truly reflects a stable steady state situation. Of course, if we started a constant volume simulation from one of the points along the S-shaped flow curve, it would slowly drift towards a stable situation of a constant volume flow curve, and the discontinuity would of course reappear. 

The observation that the non-monotonic flow curves in Fig.~\ref{fig8} do not seem to survive the large system size limit is consistent with some previous work~\cite{heussinger15}, but seems to disagree with others~\cite{mari15s,mari18} which employed however much smaller system sizes. An important conclusion from these observations is that they suggest that the underlying constitutive rheological relation for discontinuous shear-thickening are not non-monotonic, as assumed in all theoretical models~\cite{wyart,mari17,olmsted}, but more simply displays a sharp discontinuity. 

Finally, some experiments have also recently discussed non-monotonic flow curves~\cite{Sshaped}, while shear instabilities are often observed in experimental work as well~\cite{fall,poon2}. The observation of S-shaped flow curves in experiments performed with much larger systems that the ones we simulate is surprising, as we would expect these experiments to be even closer to the large $N$ limit than our simulations, and we have no explanation for this apparent discrepancy. Regarding flow instabilities, we can only reiterate the obvious statement that by construction our simulations are performed with periodic boundary conditions in perfectly homogeneous conditions, with no gradient of any kind present in the system and no boundary effects. In addition, the pressure can increase by orders of magnitude in our system and we do not need any robust machinery to maintain the sample inside the rheometer. Thus, in a sense, simulations represent an idealised experimental situation, but they clearly demonstrate that flow instabilities are not a necessary consequence of the presence of discontinuous shear-thickening. 
 
\section{Summary and conclusion}

\label{conclusions}

To summarize, we studied the overdamped Brownian dynamics of a simple model of soft repulsive spheres with frictional contacts. The model undergoes a glass transition due to thermal fluctuations, and also undergoes a jamming transition when thermal fluctuations are not present. The combination of Brownian forces and frictional forces is enough to induce a discontinuous shear-thickening behaviour. Our study has established that such behaviour could be observed over a broad range of control parameters, and can span a large range of shear stresses and shear rates, depending on the particle softness. Thus, we showed that there is no need to invoke any additional physical mechanisms or contact forces to obtain a very realistic behaviour.  

As in previous studies, the relevant microscopic mechanism is a sharp transition between frictionless and frictional rheologies occurring at large P\'eclet number but nevertheless controlled by the intensity of Brownian forces. We have exposed the relevant stress scale controlling shear-thickening in our simulations, and have demonstrated that our model is enough to reproduce a broad range of  controlled experimental studies with model colloidal particles. 

Furthermore we have carefully discussed the singular nature of the hard sphere limit in the present context, and explored the consequences of the results for experiments where the pressure (rather than the volume) is controlled. This had led us to conclude that the existence of S-shaped flow curves in our model is a finite-size effect, that is in addition not associated to any remarkable flow instability.

Finally, it should be obvious that many materials undergoing shear-thickening behaviour are not composed of model colloidal particles that we have analysed in the present work. Although our model produces flow curves that are qualitatively similar to those more complicated materials, such as cornstarch dispersions~\cite{brown,brown2010} or particles with attractive interactions~\cite{ema}, it is also clear that these particles may carry charges and have non-spherical shapes. It would be surprising that our model could reproduce their rheology at the quantitative level. This might be because in such system the shear thickening takes place via the interplay between frictional force and the electrostatic force and complex geometrical frustration. It remains to be explored how to include these more complicated features in a computational framewrok that remains tractable.

\acknowledgments
We thank H. Hayakawa, M. Otsuki, A. Ikeda, L. Hsiao, and R. Seto for valuable discussions. The research leading to these results has received funding
from the European Research Council under the European Union's Seventh
Framework Programme (FP7/2007-2013) / ERC Grant agreement No 306845, by a grant from the Simons Foundation (\#454933, Ludovic Berthier), and JSPS Kakenhi (No 15H06263, 16H04025, 16H04034, and 16H06018). 

\appendix 

\section{Derivation of the relation between the contact stiffness and the particle size}

\label{ref:laplace}

The relation $\epsilon_{\rm n} \propto a^{2}$ is explained simply using the Laplace pressure. Here, we review the corresponding derivation.
Firstly, suppose a contact force $F$ is applied to the surface of the sphere whose diameter is $a$.
Now that $F$ is balanced with the pressure difference between the inside and outside of the sphere ($\Delta P$) multiplied by the contact surface area, which reads
\be
F \sim \frac{\pi a^2 \Delta P}{4}.
\label{force2}
\ee
On the other hand, $\Delta P$ is obtained from the energy balance equation between the bulk and the surface contributions when the small perturbation ($\Delta V$) is applied such as  
\be
0= -\Delta P\Delta V +  \gamma \Delta A,
\ee  
where $\Delta V = \pi a^2 \Delta a/2 $ and $\Delta A = 2\pi a \Delta a$. When $\Delta a$ is small, the elastic energy change of the sphere is negligible because it is $\mathcal{O}((\Delta a)^2)$.
From the above relation, we obtain $\Delta P$, which reads
\be
\Delta P =\frac{4\gamma}{a}.
\label{laplace}
\ee
Using Eqs.~(\ref{force2}, \ref{laplace}) together with our normal spring contact force given by Eq.~(\ref{force}),  $F$ can be represented  as
\be
F \sim \pi \gamma a  \sim \frac{\epsilon_{\rm n}}{a}.
\ee
Accordingly we can obtain the relation
$\epsilon_{\rm n} \propto a^{2}$ which is consistent to also our findings represented in Eq.~(\ref{stiff}).
Also, substituting $\epsilon_{\rm n} \propto a^{2}$ to $\sigma_{\rm c} \approx \sigma_{\rm T}^{\rm inst} = \epsilon_{\rm n}/a^3 \sqrt{k_B T / \epsilon_{\rm n}}$, 
we obtain
\be
\sigma_{\rm c} \propto a^{-2},
\ee
which is consistent to the experimental findings~\cite{poon}.
The above argument is of course trivial when particles are real droplets, since this corresponds to the Laplace pressure. However, assuming the colloidal particles deform very little (as indeed correct for our simulated system), the argument applies also for colloids.

\end{document}